\documentclass{emulateapj}
\usepackage{amsmath,amstext}
\usepackage{wasysym} 						
\usepackage{verbatim}
\usepackage{soul}
\usepackage[T1]{fontenc}
\usepackage{apjfonts} 
\usepackage{hyperref}
\usepackage[figure,figure*]{hypcap}

\shorttitle{Modeling Turbulent Material in the CGM}
\shortauthors{Buie et al.}

\begin{document}

\title{Modeling  Photoionized Turbulent Material in the Circumgalactic Medium}

\author{Edward Buie II}
\affiliation{Arizona State University School of Earth and Space Exploration, P.O. Box 871404, Tempe, AZ 
85287, USA}
\author{William J. Gray}
\affiliation{CLASP, College of Engineering, University of Michigan, 2455 Hayward St., Ann Arbor, MI 48109, USA}
\author{Evan Scannapieco} 
\affiliation{Arizona State University School of Earth and Space Exploration, P.O. Box 871404, Tempe, AZ 
85287, USA}

\begin{abstract}
The circumgalactic medium (CGM) of nearby star-forming galaxies shows 
clear indications of \ion{O}{6} absorption accompanied by little to no \ion{N}{5} absorption. This unusual spectral signature, accompanied 
 by absorption from lower ionization state species whose columns vary by orders of magnitude along different sight lines, indicates that the CGM must be viewed as a dynamic, multiphase medium, such as occurs in the presence of turbulence. To explore this possibility, we carry out a series of chemodynamical simulations of a isotropic turbulent media, using the MAIHEM package. The simulations assume a metallicity of $0.3\ Z_{\odot}$ and a redshift zero metagalatic UV background, and they track ionizations, recombinations, and species-by-species radiative cooling for a wide range of elements. We find that turbulence with a one-dimensional (1D) velocity dispersion of $\sigma_{\rm 1D} \approx 60$ km s$^{-1}$ replicates many of the observed features within the CGM, such as  clumping of low ionization-state ions and the existence of \ion{O}{6} at moderate ionization parameters. However, unlike observations, \ion{N}{5}  often arises in our simulations with derived column densities of a similar magnitude to those of \ion{O}{6}. While higher values of $\sigma_{1D}$ lead to a thermal runaway in our isotropic simulations, this would not be the case in stratified media, and thus we speculate that more complex models of the turbulence may well match the absence of \ion{N}{5} in the CGM of star-forming galaxies.

\end{abstract}


\section{Introduction}

The circumgalactic medium (CGM) is a highly ionized plasma that resides within the dark matter halo of galaxies. It contains the reservoir of baryons for galaxy formation and has an important role in regulating galaxy evolution through inflow and outflow processes, such as galactic accretion, active galactic nuclei, and galactic winds \citep[e.g.][]{2013lilly,2015voit,2015crighton,fox2017gas,2017muratov,2017reviewARA&A..55..389T}. Its diffuse nature, however, makes direct observation difficult. 

Absorption in the spectra of background QSOs allows us to gain insight into these diffuse systems. \citet{1998steidel} summarized much of the earlier results that attempted to understand the relationship between the Intergalactic Medium (IGM) and the CGM. At the time, the newly commissioned WFPC-2 on the \textit{Hubble Space Telescope (HST)} as well as the HIRES spectrograph on the Keck telescopes, gave us the capability to observe these systems up to $z \approx 3$ with improved follow-up spectroscopy. The Keck Baryonic Structure Survey \citep[KBSS;][]{rudie2012gaseous} continued these efforts by targeting galaxies at the peak of star-formation ($z \approx 2 - 3$). It is theorized that at higher redshift there should be a high accretion rate of cold material through filamentary structures \citep[e.g.][]{birnboim2003virial,ocvirk2008bimodal,brooks2009role,faucher2011small}. They were able to map the \ion{H}{1} distribution around these systems as well as show an anti-correlation between \ion{H}{1} absorbers and impact parameter.

In recent years, the Cosmic Origins Spectrograph (COS) installed on the HST has vastly improved our sensitivity to diffuse material that absorbs in the UV \citep{2009AIPC.1135..301S}. Specifically, the COS-Halos survey allowed us to probe the CGM of $z$ $\lesssim$ 0.5, galaxies \citep{tuml2013ApJ...777...59T} by using quasar absorption-line spectroscopy. \citet{tuml2013ApJ...777...59T} probe $M_{*} = 10^{9.5} - 10^{11.5}$ \textit{M}$_{\odot}$ galaxies out to an impact parameter, $b = 150$ kpc. These observations uncovered a large amount of \ion{O}{6} absorption in the CGM of star-forming galaxies and also showed a neutral H component associated with nearly all of the galaxies.

\citet{2013werkApJS..204...17W} expanded upon this study by finding significant metal-line absorption corresponding to a cool ($T \approx 10^4-10^5$ K) CGM phase. This cooler phase is constrained to the inner CGM such that column densities and detection rates for lower ionization species decrease with increasing impact parameter. Furthermore, in \citet[][hereafter W16]{werk2016ApJ...833...54W} \ion{O}{6} absorption within the star-forming sample is further explored. They find \ion{O}{6} absorption to span the entire CGM, while also discovering \ion{N}{5} absorption to be absent in  35 of the total 38 \ion{O}{6} components.

W16 also looked at many models to explain this phenomenon; these included photoionization models such as CLOUDY \citep{2017cloudyRMxAA..53..385F}, shock ionization models \citep{2009gnatApJ...693.1514G}, collisional ionization equilibrium and non-equilibrium models \citep{2007gnatApJS..168..213G}, radiative cooling flow models \citep{2012wakkerApJ...749..157W} and more. These models either required very high ionization from an extra-galactic ultraviolet background (EUVB), unphysically long path lengths for \ion{O}{6}, or a narrow range of parameters to fit the data. 
This motivates us to explore the effects of turbulence in the CGM.

There are several CGM processes that are likely to drive significant turbulence. Theoretical work has shown that inflows should be distinctly colder than the surrounding medium \citep{ 2005coolflowMNRAS.363....2K, 2006coolflowMNRAS.368....2D, 2009coolflowApJ...700L...1K, 2011coolflowApJ...738...39S} and there is observation evidence of colder inflowing material around star-forming galaxies \citep{2012inflowApJ...747L..26R}. As this colder material flows inward, there should at least be turbulence along the boundaries between colder and hotter material. Also in the case of outflows, material, momentum, and energy are injected into the surrounding medium, which may also induce turbulence. This leads us to theorize how various ions may change in the presence of isotropic turbulence.

Here we present direct numerical simulations of a turbulent astrophysical media exposed to an EUVB, in an effort to determine  the extent to which multiphase observations of the CGM can be explained by the presence of sustained, isotropic turbulence. The paper is organized as follows: in Section 2 we outline the code used to model the CGM. In Section 3 we present our results with a focus on \ion{O}{6} and \ion{N}{5} abundances as well as compare our results to W16 and give concluding remarks.

\section{Methods}
\subsection{The MAIHEM Code}
To simulate an isotropic turbulent CGM, we use Models of Agitated and Illuminated Hindering and Emitting Media (MAIHEM\footnote{http://maihem.asu.edu/}), a three-dimensional (3D) cooling and chemistry package built using FLASH (Version 4.3), an open-source hydrodynamics code \citep{fryxell2000flash}. MAIHEM explicitly tracks the reaction network of 65 ions: including hydrogen (\ion{H}{1} and \ion{H}{2}), helium (\ion{He}{1}--\ion{He}{3}), carbon (\ion{C}{1}--\ion{C}{6}), nitrogen (\ion{N}{1}--\ion{N}{7}), oxygen (\ion{O}{1}--\ion{O}{8}), neon (\ion{Ne}{1}--\ion{Ne}{10}), sodium (\ion{Na}{1}--\ion{Na}{3}), magnesium (\ion{Mg}{1}--\ion{Mg}{4}), silicon (\ion{Si}{1}--\ion{Si}{6}), sulfur (\ion{S}{1}--\ion{S}{5}), calcium (\ion{Ca}{1}--\ion{Ca}{5}), iron (\ion{Fe}{1}--\ion{Fe}{5}), and electrons from an initial non-equilibrium state to steady state. This includes solving for dielectric and radiative recombinations, collisional ionizations with electrons, charge transfer reactions, and photoionizations by a UV background. 

This package was first developed in \citet{gray2015atomic} and later improved upon with the inclusion of an ionizing background in \citet{gray2016atomic}. Most recently, in \citet{gray2017effect}, several charge transfer reactions, radiative recombination rates, and dielectronic recombination rates from \citet{aldrovandi1973radiative,shull1982ionization,arnaud1985updated} have been added to and updated in MAIHEM. Furthermore, the cross sections for the photoionizing and photoheating rates are taken from \citet{verner1995analytic} for the inner electron shell transitions and \citet{verner1996atomic} for the outer electron shell transitions.

The equations solved by MAIHEM are given in \citet{gray2016atomic} and are invariant under the transformation $x \rightarrow \lambda x,\ t \rightarrow \lambda t,\ \rho \rightarrow \rho/\lambda$ meaning the final steady-state abundances depend only on the mean density multiplied by the driving scale of turbulence, $nL$, the one-dimensional (1D) velocity dispersion of the gas, $\sigma_{\rm 1D}$, and the ionization parameter, $U$; the ratio of number of ionizing photons to the number density of hydrogen $n_{\rm H}$, or alternatively,
\begin{equation}
\label{eq:1}
U  \equiv \frac{\Phi}{n_{\rm H} c}, 
\end{equation}
where $\Phi$ is the total photon flux of ionizing photons, and $c$ is the speed of light.

Finally, we elect to model turbulence through solenoidal modes ($\nabla \cdot F = 0$) and use an unsplit solver based on \citet{2013leeJCoPh.243..269L} to solve the hydrodynamic equations. In addition to this, we make use of a hybrid Riemann solver that utilizes the Harten Lax and van Leer (HLL) solver \citep{einfeldt1991godunov} in places with strong shocks or rarefactions and the Harten--Lax--van Leer--Contact (HLLC) solver \citep{toro1994restoration,tororiemann} in smoother flows to stabilize the code as turbulence ensues. We refer the reader to \citet{gray2015atomic} and \citet{gray2016atomic} for further details. 
\begin{figure}[h]
\includegraphics[width=1.08\linewidth]{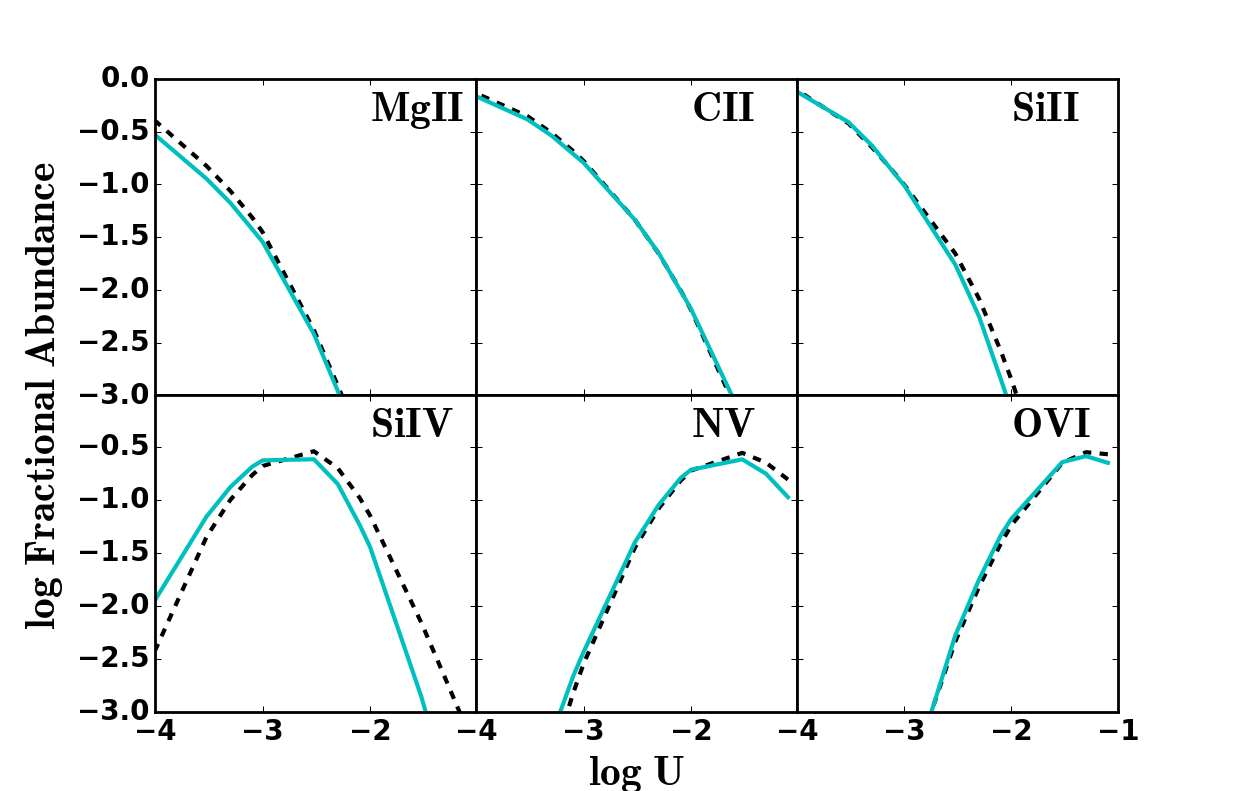}
\caption{Comparison of fractional abundances for MAIHEM (dashed black) and CLOUDY (solid cyan). MAIHEM was run with $nL_{\rm box} = 10^{21}$ cm$^{-2}$ with $\sigma_{\rm 1D} = 1$ km s$^{-1}$. CLOUDY was run with a varying hydrogen density such that the resultant $U$ was varied between $-4 <$ log $U < -1$.}
\label{fig:cloudyvflash}
\end{figure}
\subsection{Abundance Test}

MAIHEM has been rigorously tested over a wide parameter space in \citet{gray2015atomic}, \citet{gray2016atomic}. However, to confirm its accuracy under conditions similar to those in the CGM, we carry out a series of tests at a metallicity of $0.3Z_{\odot},$ which is the median value found by 
\citet{prochaska2017cos} for the CGM of low-redshift galaxies. 

MAIHEM abundances are compared to those given by CLOUDY, an open-source photoionization code \citep{ferland20132013}. Specifically, we conduct single-zone CLOUDY runs and MAIHEM runs with $nL = 10^{21}$ cm$^{-2}$ and $\sigma_{1D}$ of $1$ km s$^{-1}$ that both use the \citet[][hereafter HM2012]{2012ApJ...746..125H} EUVB. We also omit the presence of molecules in our CLOUDY tests. Here we show some of the ion abundance comparison plots in Figure \ref{fig:cloudyvflash} for ions that we will focus on throughout the paper. Ion abundances for MAIHEM are found using solar relative abundances from \citet{2010lodderASSP...16..379L}.

We note that out of the abundances plotted, all of them closely agree with CLOUDY with the exception of \ion{Si}{4} at log $U > -2$. It seems that MAIHEM slightly overpredicts the higher ion fractional abundance, which takes away from the fractional abundance at high $U$.

\subsection{Model Parameters}
\begin{figure*}[t]
\hspace*{-0.4in}
\centering
\includegraphics[width=1.13\linewidth]{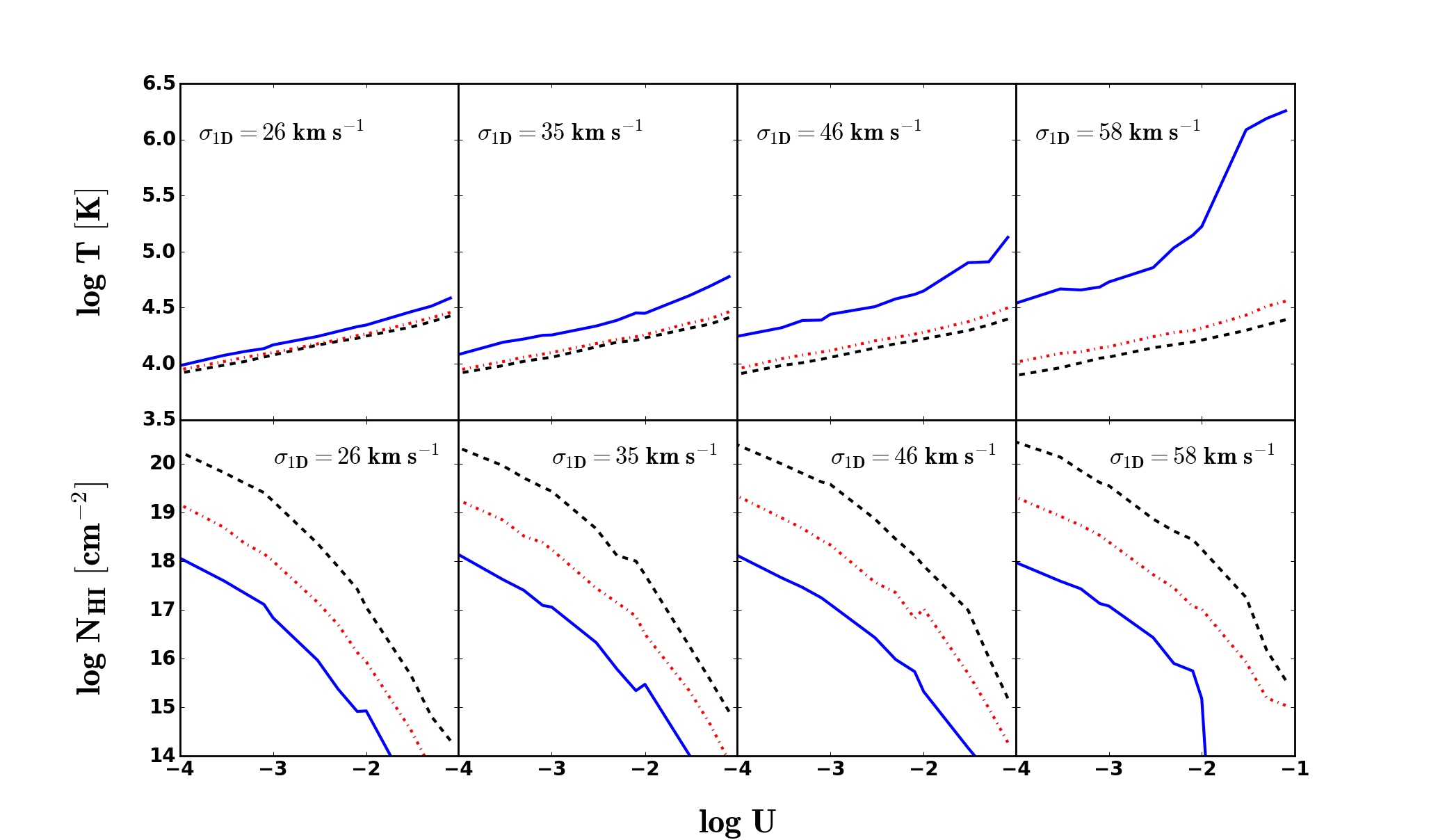}
\caption{log Temperature vs. log $U$ (top row) and log $N_{\rm H\ 1}$ vs. log $U$ (bottom row) for the $nL = 10^{19}$ (solid blue), $10^{20}$ (dotted-dashed red), and $10^{21}$ (dashed black) cm$^{-2}$ runs. From left to right, $\sigma_{\rm 1D} = 26,\ 35,\ 46,\ 58$ km s$^{-1}$.}
\label{fig: temp_H1}
\end{figure*}
Following this abundance test, we conduct a suite of simulations to see the effect of isotropic turbulence on ion abundances. Our simulations force turbulence by stirring via solenoidal modes \citep{pan2010} with wavenumbers that vary between $1 \leqslant L_{\rm box} \lvert k\rvert/2\pi \leqslant 3.$ This ensures that the average driving scale of turbulence is $k^{-1} \simeq 2L_{box}/2\pi$ where $L_{\rm box}$ is set to 100 pc. 

We carry out these simulations in a 128$^{3}$ periodic box and begin with a uniform density of $n = 10^{-2} - 1$ cm$^{-3}$, with the lower end corresponding to the density of the CGM \citep{2017reviewARA&A..55..389T}. These densities correspond to a driving scale of turbulence that varies from $nL_{\rm box} = 10^{19} - 10^{21}$ cm$^{-2}$. The medium is initialized with a fractional ion abundance that corresponds to collisional ionization equilibrium at a temperature of $T = 10^{5}$ K. 

The range of $\sigma_{\rm 1D}$ that we test is varied between 26 and 60 km s$^{-1}$ following estimates of the nonthermal velocity components of line-widths measurements (W16). The box is irradiated with a redshift zero HM2012 EUVB whose strength is quantified by $U$, such that it varies between $-4 < $ log $U < -1$ in a single run by progressing in increments of log $U \sim 0.2$ once a steady state is reached. To determine when the box has reached a steady state, the average global abundances were calculated every 10 time steps. To ensure that our simulations reached a steady state, we impose a cutoff value to the change in ion abundances of 0.03 to prevent ions with small abundances from stopping the progression of the ionization parameter. This change in fractional abundances is found with
\begin{equation}
\frac{\Delta X_{i}}{X_{i}} = \frac{\overline{X_i^a} - \overline{X_i^b}}{\overline{X_i^a}},
\end{equation}
where $X_{i}$ is the abundance of ion $i$ and $X_i^a$ and $X_i^b$ are the averaged ion abundances. Only when all fractional ion abundances are below this cutoff value do we progress to higher $U$.  We vary the background with this range of $U$ to match observational data from the COS-Halos survey \citep{werk2014cos}.

\section{Results}

\begin{figure*}
\centering
\includegraphics[width=1.0\linewidth,height=1\textheight]{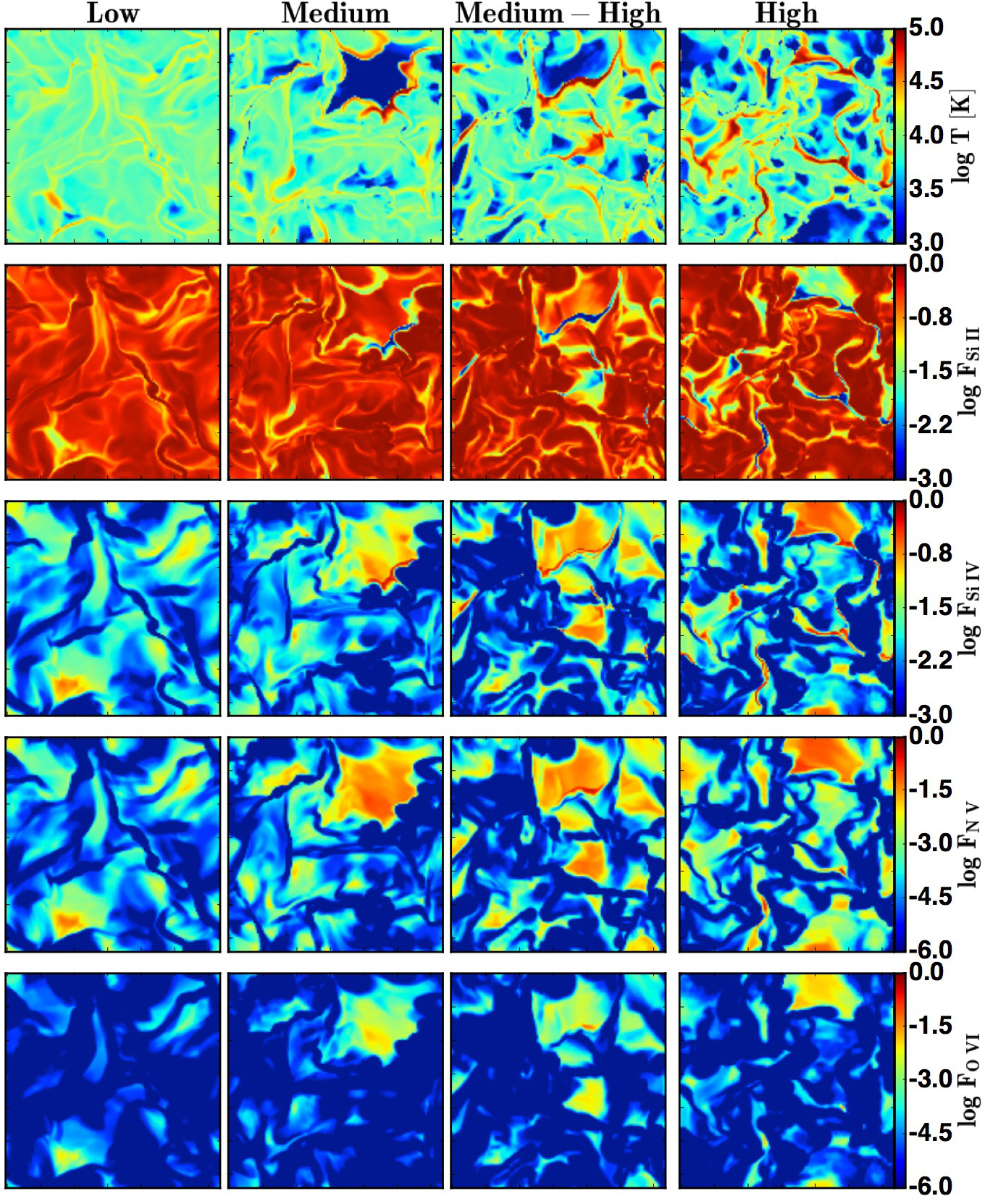}
\caption{Temperature (first row), \ion{Si}{2} (second row), \ion{Si}{4} (third row), \ion{N}{5} (fourth row), and \ion{O}{6} (bottom row) slices for $\sigma_{\rm 1D} = 26$ (Low), $35$ (Medium), $46$ (Medium-High), and $58$ (High) km s$^{-1}$ at log $U =\ -4$. Each column gives slices from the aforementioned turbulence run.}
\label{fig:u4slice}
\end{figure*}

\begin{figure*}
\centering
\includegraphics[width=1.0\linewidth,height=1\textheight]{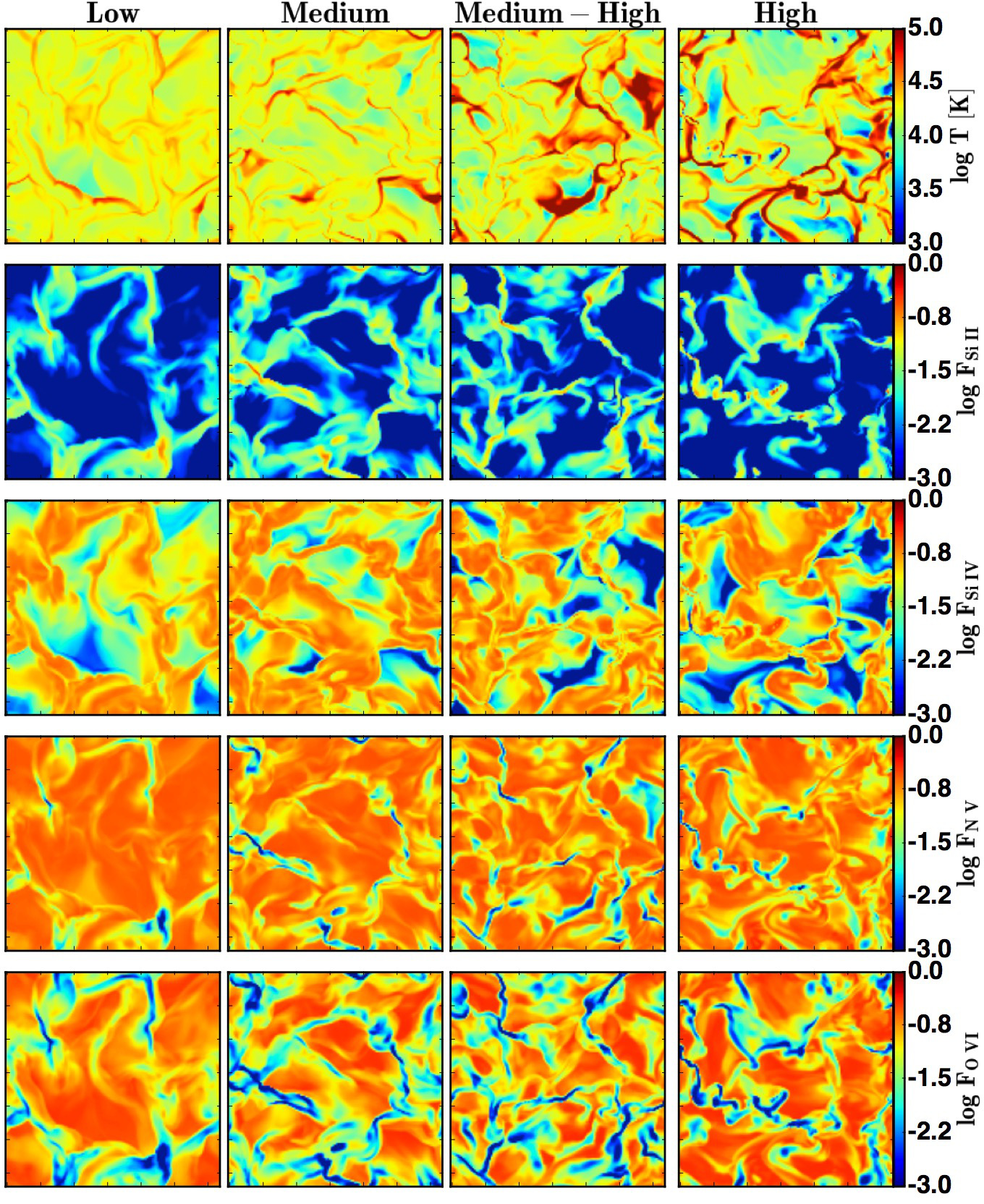}
\caption{Temperature (first row), \ion{Si}{2} (second row), \ion{Si}{4} (third row), \ion{N}{5} (fourth row), and \ion{O}{6} (bottom row) slices for $\sigma_{\rm 1D} = 26$ (Low), $35$ (Medium), $46$ (Medium-High), and $58$ (High) km s$^{-1}$ at log $U =\ -2$. Each column gives slices from the aforementioned turbulence run.}
\label{fig:u2slice}
\end{figure*}

\begin{figure*}
\hspace*{-0.4in}
\includegraphics[width=1.1\linewidth]{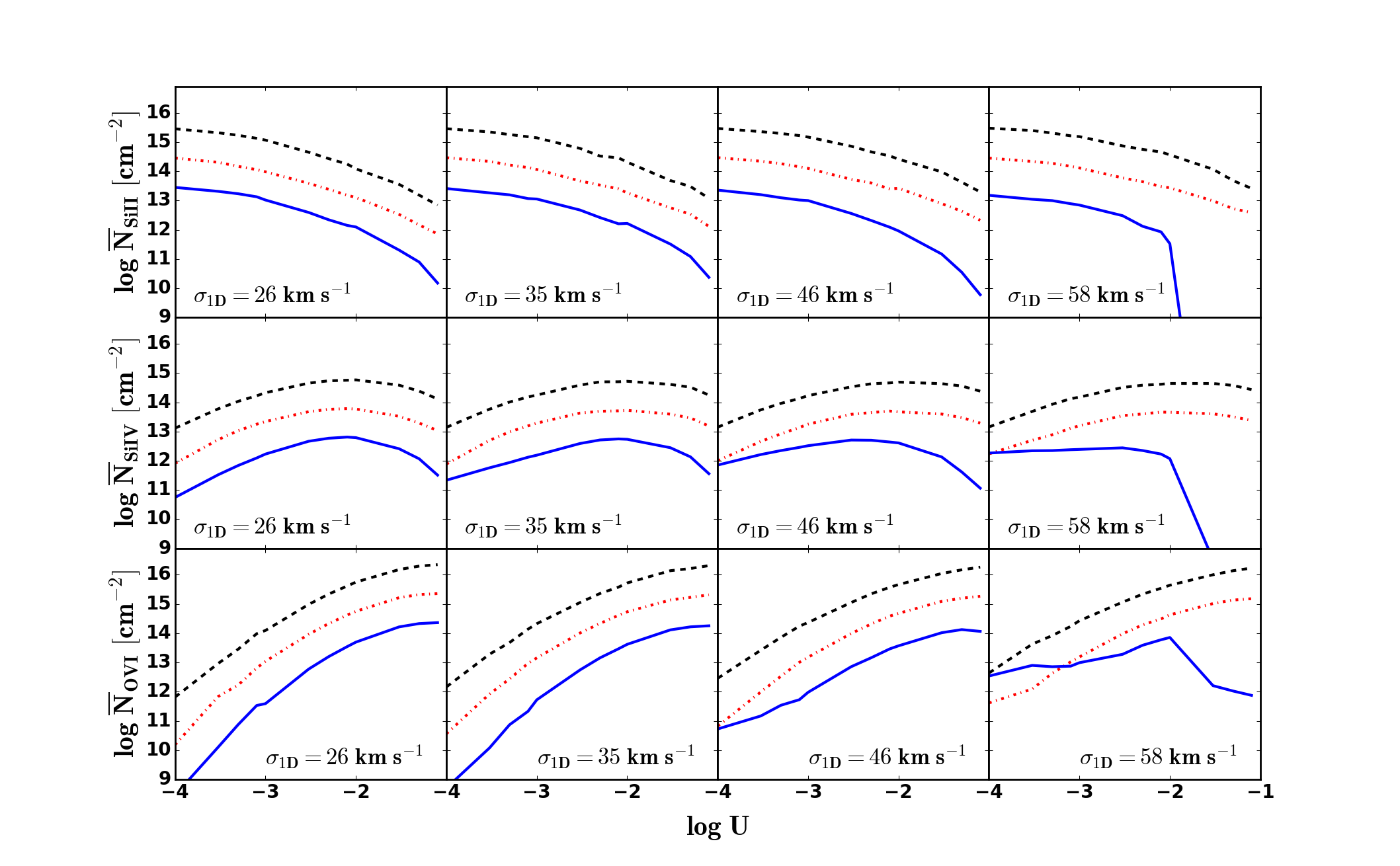}
\vspace*{-0.2in}
\caption{log $N_{\rm Si\ II}$  (first row), log $N_{\rm Si\ IV}$  (second row), and log $N_{\rm O\ VI}$  (third row) vs log $U$ for the $nL=10^{19}$ (solid blue), $10^{20}$ (dotted-dashed red), and $10^{21}$ (dashed black) cm$^{-2}$ runs. From left to right, $\sigma_{\rm 1D} = 26,\ 35,\ 46,\ 58$ km s$^{-1}$.}
\label{fig:si2_si4_o6}
\end{figure*}

\subsection{General Features}

The CGM is traced by a cool $T \lesssim 10^{5}$ K phase along with a hotter $T \gtrsim 10^{6}$ component \citep{tuml2013ApJ...777...59T,werk2014cos,werk2016ApJ...833...54W,prochaska2017cos}. The top row of Figure \ref{fig: temp_H1} shows the density-weighted temperature for all of our runs from left to right, $\sigma_{\rm 1D} = 26,\ 35,\ 46,\ 58$ km s$^{-1}$. We find that this density-weighted temperature falls within the range of temperatures observed in the CGM of nearby galaxies. We also look to match the large H I component observed in these systems and show the H I column density, $N_{\rm H\ 1}$, in the bottom row of Figure \ref{fig: temp_H1}; finding that our results match the log $N_{\rm H\ 1} = 14 - 20$ cm$^{-2}$ range found in \citet{tuml2013ApJ...777...59T}.

Figures \ref{fig:u4slice} and \ref{fig:u2slice} show temperature and ionic distributions from
$nL = 10^{20}$ cm$^{-2}$ simulations with log $U = -4$ and $-2,$ respectively.  Here, \ion{Si}{2} and \ion{Si}{4}, which are commonly observed in the halos of low-redshift galaxies \citep{2013werkApJS..204...17W,2016borthakurApJ...833..259B,2017heckman}, are chosen to sample low and intermediate ionization state material, respectively. 
\ion{O}{6}, on the other hand, serves to sample the higher ionization state material, commonly observed in the halos surrounding $L_{*}$ galaxies \citep[W16,][]{oppen2016MNRAS.460.2157O}, and likely arising from interactions between different phases as caused by processes that inject energy  and material into the CGM \citep[e.g.][]{2013hummels,2013shen,2016ford,2017gutcke,2018hani,2018liang,2018nelson}. Finally, \ion{N}{5}, which has an ionization potential slightly below \ion{O}{6}, is chosen because it is nearly absent in the CGM of star-forming galaxies for reasons that remain to be determined. 

At lower turbulent velocities ($\sigma_{\rm 1D} < 26$ km s$^{-1}$), the gas is subsonic, yielding weak shocks and small temperature gradients, which ensures that the medium is dominated by a single ionization state material. Low ions, defined to be singly ionized metals, dominate if log $U$ = -4 while intermediate ions dominate in the log $U$ = -2 case. We also begin to see larger temperature gradients as this turbulence is increased, caused by stronger shocks, which sweep across cells carrying newly ionized material. An example of this appears in the upper-right corner of the Medium-High run in Figure \ref{fig:u4slice}, which features a downward shock that leaves \ion{N}{5} and \ion{O}{6} in its wake. 

\begin{figure*}[t]
\hspace*{-0.4in}
\includegraphics[width=1.1\linewidth]{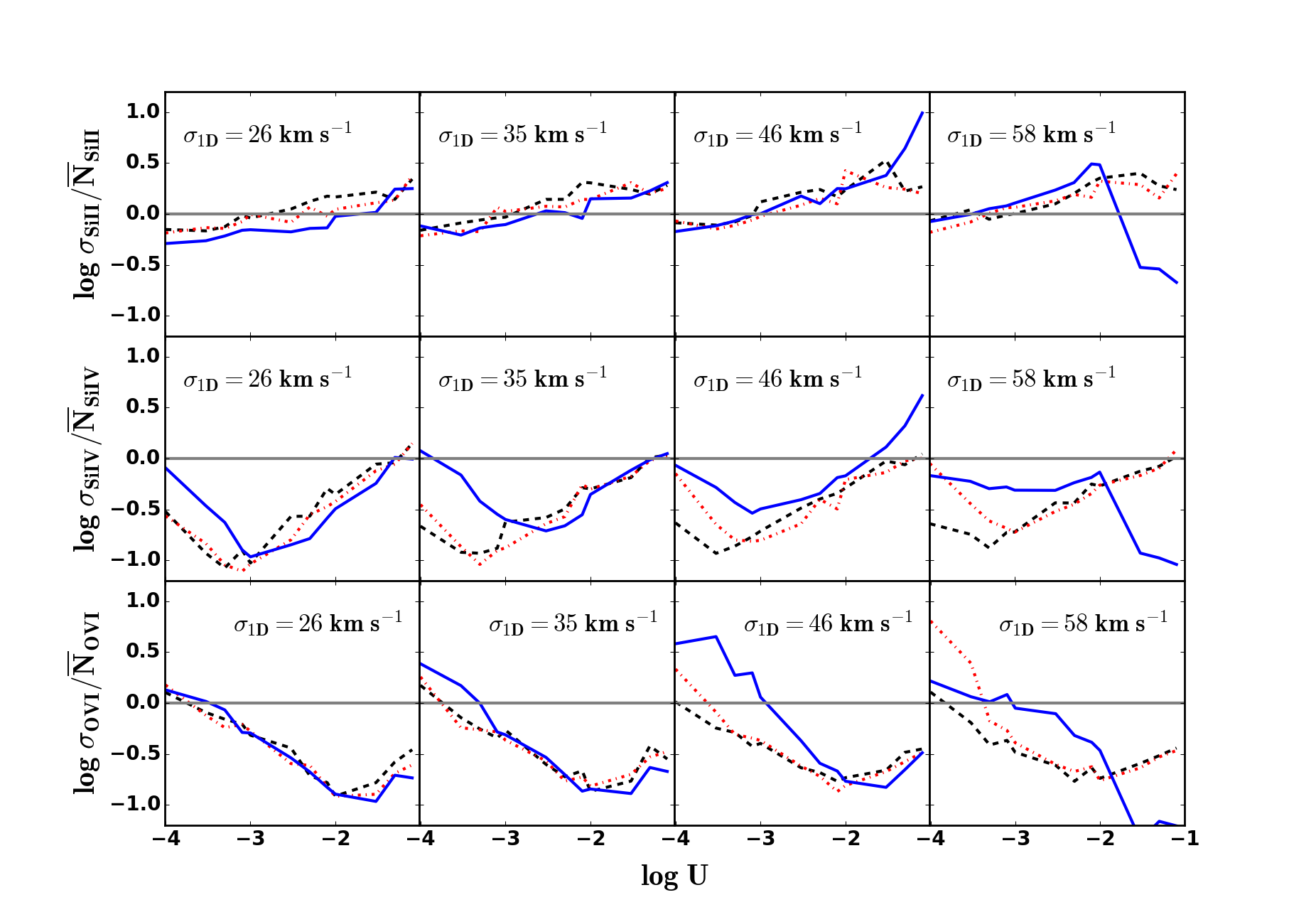}
\vspace*{-0.2in}
\caption{log $\sigma_{\rm SiII}/\overline{N}_{\rm Si II}$  (first row), log $\sigma_{\rm SiIV}/\overline{N}_{\rm Si IV}$  (second row), and log $\sigma_{\rm O VI}/\overline{N}_{\rm O VI}$ (third row) vs. log $U$,  for $\sigma_{\rm 1D} = 26$, $35$, $46$, and $58$ km s$^{-1}$ (from left to right). Normalized standard deviations from the $nL = 10^{19}$ cm$^{-2}$ are plotted in solid blue, $nL = 10^{20}$ cm$^{-2}$ in dotted-dashed red, and $nL = 10^{21}$ cm$^{-2}$ in dashed black. The solid gray line is shown to indicate a normalized standard deviation of one. }
\label{fig:low_inter}
\end{figure*}

\subsection{Low and Intermediate Ionization State Ions}

The mean of their respective column densities of \ion{Si}{2}, \ion{Si}{4}, and \ion{O}{6} are shown in Figure \ref{fig:si2_si4_o6}.
As mentioned previously, the turbulent state of the medium and its background determine the dominant ionization state. However, sight lines of \ion{N}{5} and \ion{O}{6} in the lower background case, and \ion{Si}{2} remains in the higher background case with mass fractions of a few percent, which correspond to column densities of $N_{\rm col} \lesssim 10^{14}$ cm$^{-2}$. Observationally, low ionization-state ions tend to be more clumpy \citep{2014pieri} and 
 congregate in the inner halo of a galaxy 
\citep{2014liang} while more highly ionized metals are more abundant in the outer CGM \citep[W16;][]{2015johnson,oppen2016MNRAS.460.2157O,2017oppenheimer}. This translates to the Low and Medium runs being more comparable to conditions of the inner CGM while the Medium-High and High runs are more representative of the outer CGM.

\citet{2013werkApJS..204...17W} find low ionization-state ions (\ion{C}{2}, \ion{Si}{2}, \ion{Mg}{2}, etc.) to have column densities that vary by orders of magnitude, in contrast to the less variable \ion{O}{6} column densities, suggesting a patchy CGM. This is seen in the Figure \ref{fig:u2slice}, as turbulence causes the low ionization-state ions, as traced by \ion{Si}{2},  to congregate into clumps.

To measure the `clumpiness' of low ionization-state ions, we computed a normalized standard deviation of the \ion{Si}{2}, \ion{Si}{4}, and \ion{O}{6} column densities for all of our runs shown in Figure \ref{fig:low_inter}. We find $N_{\rm SiII}$ to have the highest average normalized standard deviation, while also having the weakest dependence on the mean density of the medium parameterized by $n L$. We further point out that this patchiness evolves as the turbulence is increased; such that higher turbulent velocities tend to move the low ionization-state ions into denser clumps.

The intermediate and higher ionization-state ions also tend to evolve as the turbulence is increased, and they depend more on the strength of the background radiation field. Specifically, \ion{Si}{4} has a minimum normalized standard deviation at lower $U$, at values at which the total \ion{Si}{4} column density is highest. As the ionizing background increases, these intermediate ions congregate into filaments, as shown in Figure \ref{fig:u2slice}, showing more patchy behavior. Finally, the normalized standard deviation of \ion{O}{6} decreases monotonically as a function of increasing turbulence.

\begin{figure*}[t]
\hspace*{-0.4in}
\centering
\includegraphics[width=1.13\textwidth]{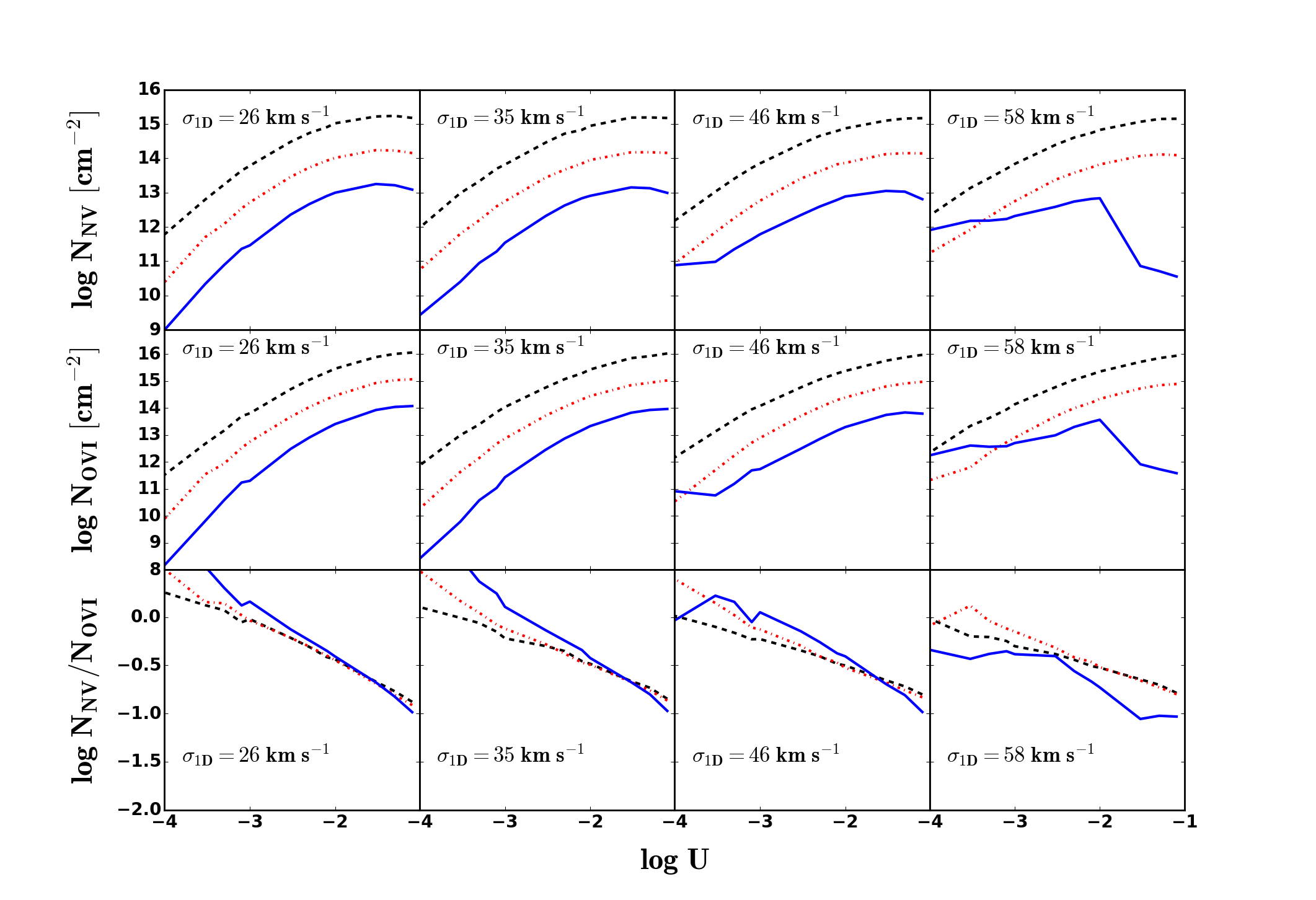}
\caption{log $N_{\rm N\ V}$ vs. log $U$ (top row), log $N_{\rm O\ VI}$ vs. log $U$ (middle row), and log $N_{\rm N\ V}/N_{\rm O\ VI}$ vs. log $U$ (bottom row) for the $nL = 10^{19}$ (solid blue), $10^{20}$ (dotted-dashed red), and $10^{21}$ (dashed black) cm$^{-2}$ runs. From left to right, $\sigma_{\rm 1D} = 26,\ 35,\ 46,\ 58$ km s$^{-1}$.}
\label{fig:N5_O6_no}
\end{figure*}

\subsection{\ion{N}{5} and \ion{O}{6}} 

In Figure \ref{fig:N5_O6_no} we show the average \ion{N}{5} column density, $N_{\rm N\ V}$ (top row), \ion{O}{6} column density, $N_{\rm O\ VI}$ (middle row), and the ratio of $N_{\rm N\ V}$ to $N_{\rm O\ VI}$ (bottom row) as a function of the ionization parameter. Here we see that the \ion{N}{5} column densities tend to be slightly higher than those of \ion{O}{6} at low $U$. This trend continues until a higher turbulent state is reached ($\sigma_{\rm 1D} \geqslant 46$ km s$^{-1}$), at which point $N_{\rm O\ VI} \geqslant N_{\rm N\ V}$ at the lowest $U$. We also see that $N_{\rm N\ V}$ seems to increase at a slower rate than $N_{\rm O\ VI}$ allowing for a lower $N_{\rm N\ V}$ to $N_{\rm O\ VI}$ ratio at higher $U$. 

As the turbulence approaches supersonic speeds, both the minimal $N_{\rm N\ V}$ and $N_{\rm O\ VI}$ increase for each run while eventually reaching a maximum at the highest $U$. However, at $\sigma_{\rm 1D} \geqslant 58$ km s$^{-1}$, the $nL = 10^{19}$ cm$^{-2}$ run experiences a peak in \ion{N}{5} and \ion{O}{6} at log $U \approx -2$. Following this, these ions undergo a sharp decline in abundance. Turbulence seems to lower both the $N_{\rm N\ V}$ and $N_{\rm O\ VI}$ gradients across $U$ allowing for larger column densities to be found at lower $U$. 

The overall density of the gas sampled is important in determining the ionization state and $N_{\rm O\ VI}$. This becomes even more important as the medium becomes more turbulent. W16 find log $N_{\rm O\ VI}$ to have a median value of $10^{14.5}$ cm$^{-2}$ for nearby star-forming galaxies. Looking at the $\sigma_{\rm 1D} = 58$ km s$^{-1}$ plot in the middle row of Figure \ref{fig:N5_O6_no}, we see that depending on the total density of the medium, an \ion{O}{6} column density of $10^{14.5}$ cm$^{-2}$ may be found anywhere between $-3 <$ log $U < -2$.

We continue to compare our results to those found in W16 by plotting the column density ratio \ion{N}{5}/\ion{O}{6} as a function $U$ in Figure \ref{fig:N5_O6_no}. As the  turbulence increases, this ratio tends to flatten while never reaching 0.1. However, W16 find their sample to have a median value of $\approx 0.1$ while noting most of their ratios to be upper bounds due to the non-detection of \ion{N}{5} in the bulk of associated \ion{O}{6} components. 
\begin{figure*}
\hspace*{0in}
\includegraphics[width=1.04\linewidth]{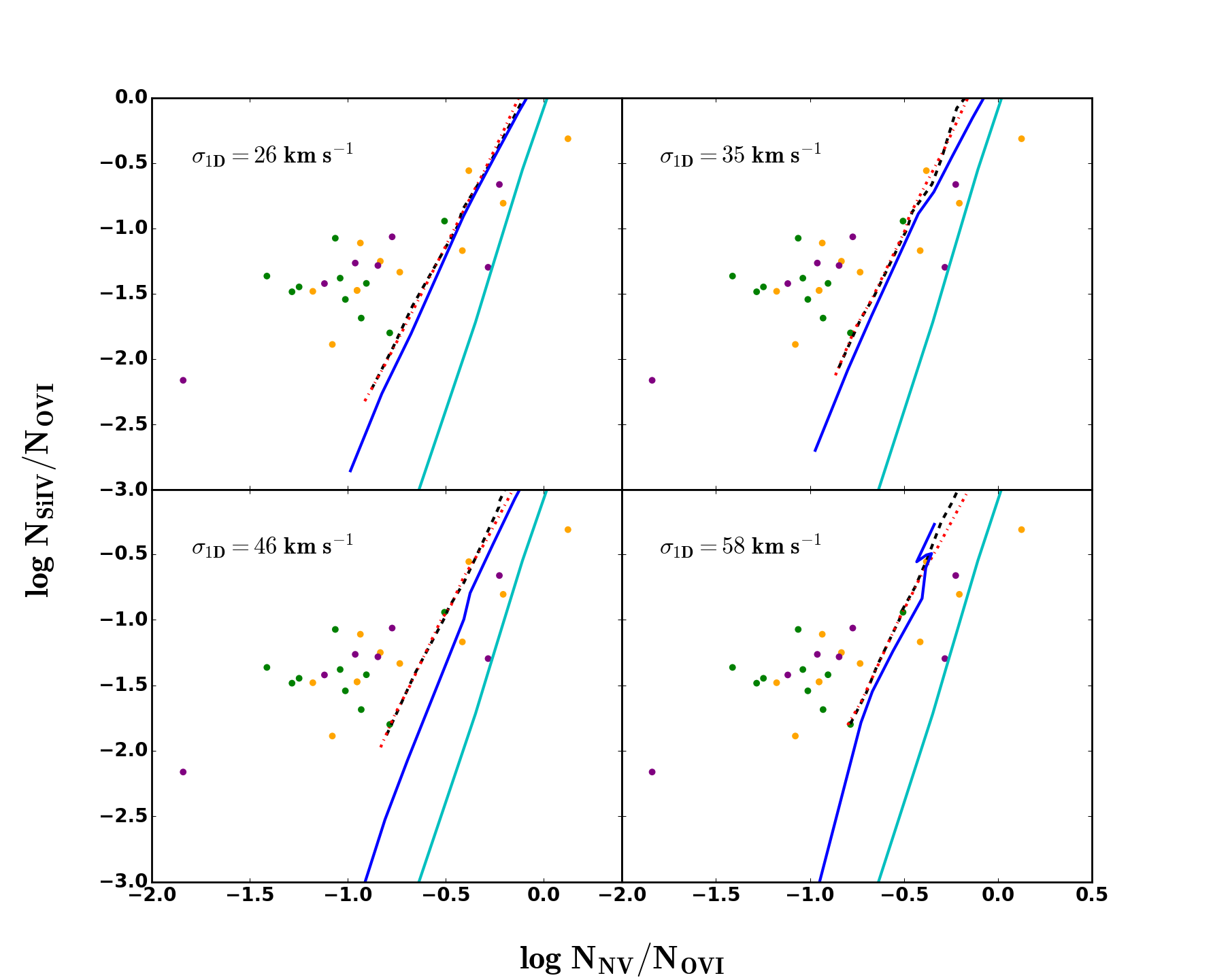}
\caption{log $N_{\rm Si\ IV}/N_{\rm O\ VI}$ vs log $N_{\rm N\ V}/N_{\rm O\ VI}$ for the $nL = 10^{19}$ (solid blue), $10^{20}$ (red dotted-dashed), and $10^{21}$ (black-dashed) cm$^{-2}$ runs. From left to right, $\sigma_{\rm 1D} = 26,\ 35$ km s$^{-1}$ (top row) and $46,\ 58$ km s$^{-1}$ (bottom row). These are overlaid on the matched \ion{Si}{4}, \ion{N}{5}, and \ion{O}{6} components from W16 Figure 12. Green circles represent the ``broad'' type \ion{O}{6} absorption, orange circles are ``narrow'' type absorbers, and purple circles are ``no-low'' type absorbers. As a reference, we plot a single-zone CLOUDY model photoionized by an HM2012 EUVB (solid cyan) with -4 < log $U$ < -1.}
\label{fig:sion}
\end{figure*}

In Figure \ref{fig:sion}, we replicate Figure 12 in W16 by plotting the log $N_{\rm Si\ IV}/N_{\rm O\ VI}$ versus log $N_{\rm N\ V}/N_{\rm O\ VI}$. We have overlaid our derived ratios with matched \ion{Si}{4}, \ion{N}{5}, and \ion{O}{6} components from W16. Here we see that turbulence moves this curve slightly leftward toward the data but not nearly enough to match the majority of the data, which require much lower levels of \ion{N}{5}.

The nonnegligible amount of \ion{N}{5} present in isotropic turbulence  may be due to two reasons: electron collisional ionization from the shocks, and the constraints imposed by isotropy. 

The temperature associated with a shock within the high turbulence run is $\approx 10^5\ {\rm K} \approx 10\ {\rm eV}$. At this temperature, the recombination rate coefficient for the primary \ion{N}{5} to \ion{N}{4} reaction, $\rm N^{4+} + \textit{e}^- = N^{3+}$, is much lower than  the primary \ion{O}{6} to \ion{O}{5} reaction, $\rm O^{5+} + \textit{e}^- = O^{4+}$. This means that \ion{N}{5} survives longer than \ion{O}{6}. 

However, as the ionization potentials of \ion{N}{4} and  \ion{O}{5} are 77 and  114 eV, respectively, even in the high turbulence case it is only the strongest shocks that are able to produce \ion{N}{5} and  \ion{O}{6}, with \ion{N}{5} being produced more efficiently than \ion{O}{6}. On the other hand, the ionization potential of \ion{N}{5} is only 98 eV, while the ionization potential of \ion{O}{6} is 138 eV.  This means it is easier to covert  \ion{N}{5} into \ion{N}{6} than it is to convert \ion{O}{6} to \ion{O}{7} and that the small value of $N_{\rm N\ V}/N_{\rm O\ VI}$ observed in star-forming galaxy halos must be caused by the production of \ion{N}{6}. However, this would require even stronger shocks, which are not sustainable in an isotropic environment.

Instead, moving to isotropic turbulence levels above $\sigma_{\rm 1D} \approx 60$ km s$^{-1}$ causes a thermal runaway.  This is because the strongest shocks in such a situation produce ions that cool much less efficiently than those found at $\approx 10^5K,$ and this inefficiency of cooling increases at higher temperatures. Thus, no steady-state condition can be achieved \citep{gray2015atomic,gray2016atomic}.

Note however that such levels of continuous turbulence are compatible with an overall steady state in an anisotropic environment, as could occur if turbulence occurred within strongly stratified medium.  In this case, thermal runaway may be avoided as hot material cycles convectively upwards in the gravitational potential while a region of efficient cooing remains near the center \citep[e.g.][]{scannapieco2012,sur2016}. This remains a promising possibility for explaining the $N_{\rm N\ V}/N_{\rm O\ VI}$ observations, but it requires spanning a considerably larger parameter space than the general isotropic case considered in this study.

\section{Discussion and Summary}
Turbulence may be present in the CGM due to outflowing and inflowing processes that inject energy, momentum, and material into the surrounding medium. Recent observations of the CGM's surrounding nearby star-forming galaxies find low ionization-state ions (\ion{C}{2}, \ion{Si}{2}, \ion{Mg}{2}, etc.) with column densities that vary by orders of magnitude, substantial \ion{O}{6} absorption, and very small amounts of \ion{N}{5}. This motivated us to simulate an isotropic turbulent medium exposed to an HM2012 EUVB to see how ion abundances change under these conditions and whether isotropic turbulence can resolve this \ion{N}{5} and \ion{O}{6} anomaly.

To test the effect of isotropic turbulence on ion abundances, we have conducted a suite of numerical simulations of a turbulent astrophysical medium exposed to an extra-galactic background using MAIHEM. We first test MAIHEM in CGM-like conditions and find nice agreement between these tests and CLOUDY results.

We then detail our results, finding that turbulence is effective at producing density and temperature gradients that span nearly two orders of magnitude. Using \ion{Si}{2}, \ion{Si}{4}, and \ion{O}{6}, as samples for low, intermediate, and high ions, respectively, we show that a moderate amount of turbulence ($\sigma_{\rm 1D} = 26$ km s$^{-1}$) is able to produce a medium dominated by multiple ionization states. In these runs, subdominant species still maintain sight lines with fractional abundances at the percent level, low ionization-state species are congregated into clumps, and medium and high ionization-state species are more widespread.

We go on to investigate \ion{O}{6} and \ion{N}{5} as well as compare our results to those in W16 finding that although we are able to produce larger \ion{O}{6} column densities at lower $U$, we also produce \ion{N}{5} at a level comparable to that of \ion{O}{6}, showing that isotropic turbulence is not able to match all observations of nearby star-forming CGM.

We speculate on this and conclude it to be a consequence of the shocks present within higher turbulent runs, which causes a thermal runaway if the velocity dispersion is increased above the values reported here. This is partially a consequence of the assumption of an isotropic medium, which causes shocks and the energy they carry to be spread throughout the entire volume.

We also look at the recombination and ionization rate coefficients for the relevant reactions and find \ion{N}{5} is produced faster and lives longer than \ion{O}{6}. In addition to this, \ion{N}{5} is ionized faster than \ion{O}{5} leading us to conclude that \ion{N}{5} present in the CGM of star-forming galaxies is able to quickly ionize to \ion{N}{6} leaving little trace of \ion{N}{5}.

In a future paper, we will modify MAIHEM with the addition of gravity to simulate a stratified medium. We predict that this will allow for more effective cooling at higher velocity dispersions, a situation that may be important in resolving the origin of the discrepancy between \ion{O}{6} and \ion{N}{5}. 

We would like to thank Jessica Werk for providing the data plotted in Figure  \ref{fig:sion} as well as Sanchayeeta Borthakur for her advice on absorption lines measures of low-redshift galaxy halos. E.B.II was supported by the National Science Foundation Graduate Research Fellowship Program under grant No. 026257-001. E.S. gratefully acknowledges the Simons Foundation for funding the workshop Galactic Winds: Beyond Phenomenology, which helped to inspire this work. He was supported by the NSF under grant AST14-07835 and NASA theory grant NNX15AK82G.  Simulations presented in this work were carried out  on the NASA PLEIADES supercomputer maintained by the Science Mission Directorate High-End Computing program and on the Stampede2 supercomputer at Texas Advanced Computing Center (TACC) through Extreme Science and Engineering Discovery Environment (XSEDE) resources under grant TGAST130021.

\textit{Software}: FLASH \citep[v4.3]{fryxell2000flash}, CLOUDY \citep{ferland20132013}, yt \citep{2011turk}

\bibliographystyle{yahapj}
\bibliography{references}

\end{document}